\documentclass{article}

\usepackage{arxiv}

\usepackage[utf8]{inputenc} 
\usepackage[T1]{fontenc}    
\usepackage{hyperref}       
\usepackage{url}            
\usepackage{booktabs}       
\usepackage{amsfonts}       
\usepackage{nicefrac}       
\usepackage{microtype}      
\usepackage{lipsum}
\usepackage{graphicx}
\usepackage{natbib}
\graphicspath{ {./images/} }
\usepackage{url,hyperref,lineno,microtype,subcaption}
\usepackage[onehalfspacing]{setspace}
\usepackage{pdflscape}
\usepackage[table,xcdraw]{xcolor}

\title{Vulnerability analysis in Complex
Networks under a Flood Risk Reduction
point of view}

\author{Leonardo B. L. Santos \footnotemark[1]\thanks{Centro Nacional de Monitoramento e Alertas de Desastres Naturais (Cemaden), Brazil}, \footnotemark[2]\thanks{Humboldt University of Berlin, Germany}\\
leonardo.santos@cemaden.gov.br
\And
Giovanni G. Soares \footnotemark[3]\thanks{Instituto Nacional de Pesquisas Espaciais (INPE), Brazil}\And
Tanishq Garg \footnotemark[4]\thanks{Indian Institute of Technology Kharagpur, India}\And
Aurelienne A. S. Jorge \footnotemark[3]\And
Luciana R. Londe\footnotemark[1]\And
Regina T. Reani\footnotemark[1]\And
Roberta B. Bacelar\footnotemark[5]\thanks{Anhanguera College, Brazil}\And
Carlos E. S. Oliveira\footnotemark[6]\thanks{Federal University of Ouro Preto, Brazil}
\And
Vander L. S. Freitas\footnotemark[6]\And
Igor M. Sokolov\footnotemark[2]}

\begin{document}
\maketitle
\begin{abstract}


The measurement and mapping of transportation network vulnerability to natural hazards constitute subjects of global interest for a sustainable development agenda and as means of adaptation to climate changes. During a flood, some elements of a transportation network can be affected, causing loss of lives. Furthermore, impacts include damage to vehicles, streets/roads, and other logistics services - sometimes with severe economic consequences. The Network Science approach may offer a valuable perspective considering one type of vulnerability related to network-type critical infrastructures: the topological vulnerability. The topological vulnerability index associated with an element is defined as the reduction on the network’s average efficiency due to the removal of the set of edges related to that element. In this paper we present the results of a systematic literature overview and a case study applying the topological vulnerability index for the highways in the state of Santa Catarina (Brazil). We produce a map considering that index and the areas susceptible to urban floods and landslides. The risk knowledge, combining hazard and vulnerability, is the first pillar of an Early Warning System, and represent an important tool for stakeholders from the transportation sector in a disaster risk reduction agenda.

\end{abstract}
\newpage
\section{Introduction}

In a scenario of global change, some climatic and extreme weather events are expected to increase in frequency and intensity and cause more social and economic impacts in several sectors, such as transportation systems and urban mobility. As presented in several papers in literature, the cost for repairing transport assets after either an urban flood or landslide represents a significant percentage of the total damage cost of several recent disasters around the world (\cite{Pregnolato2017, Koks2019, Randil2022, Marulanda2022}) and in Brazil (\cite{Santos2017a, Parizzi2022, Zakhia2021}).

To mitigate those impacts, it is necessary to evaluate the risk associated with disasters and the best ways to deal with them. ``Disaster risk reduction is aimed at preventing new and reducing existing disaster risk and managing residual risk, all of which contribute to strengthening resilience and therefore to the achievement of sustainable development'' (\cite{UNISDR2017}).

For disaster risk reduction, vulnerability is a key concept. There are several types and meanings for vulnerability. According to \cite{Wisner1994}, vulnerability represents ``the characteristics of a person or group and their situation that influence their capacity to anticipate, cope with, resist and recover from the impact of a natural hazard (an extreme natural event or process)''. The UN Office for disaster risk reduction also includes assets and systems as subjects to vulnerability: ``The conditions determined by physical, social, economic and environmental factors or processes which increase the susceptibility of an individual, a community, assets or systems to the impacts of hazards'' (\cite{UNISDR2017}).

In the transportation systems' literature, there are also different meanings for vulnerability (\cite{Schlogl2019}). For example, \cite{Berdica2002} suggested that network vulnerability should be understood as ``susceptibility to incidents that can result in considerable reductions in road network serviceability'', and \cite{Taylor2006} understood network vulnerability as the extent of a failure to impact the original purpose of the system.

Vulnerability is a key idea in Network Science as well. Due to its generality for representing the system topology (relation among the elements on the system), Network Science approaches have been applied to a huge number of very different areas (\cite{Newman2010}). Section \ref{sec:slm} presents a Systematic Literature Mapping of papers related to road networks, natural disasters and network vulnerability. The listed works are predominantly from 2020, published in interdisciplinary and transport journals. The most frequent type of disaster and data origin are flooding/landsliding and USA/China.

This paper presents a formulation for a vulnerability index based on efficiencies of the system of networks. It aims to locate the most vulnerable links in a transportation network and to assess whether these links are susceptible to hazards and disruptions. The idea is presented as a case study on a set of highways, which are mapped based on vulnerability index and disaster susceptibility data.

\section{Material and Methods}

\subsection{Study area}

Brazil is among the ten countries most affected by weather-related disasters in the last 20 years (\cite{UNISDR2017}). Santa Catarina state, located in the Brazilian Southern region, is particularly affected by disasters - there is an annual mean of 64 damage records triggered by hydrological processes, such as floods in Santa Catarina municipalities (\cite{Herrmann2014}). The maximum value was achieved in 2008, when the material losses summed almost 1 billion US Dollar (\cite{Herrmann2014}). 



According to the last census track (2010), there are 295 municipalities and more than 6 million inhabitants in the state. The State's HDI - Human Development Index - is 0.774 and it is the third in the Brazilian HDI ranking (\cite{IBGE2010}). Despite the high socio-economic indicators for municipalities from Santa Catarina state, there are many communities at risk in those places due to characteristics of land occupation (\cite{Londe2014, londe2015impactos}). The mountainous relief in the east side determined the human settlement in the fluvial plains, which are areas naturally prone to floods. Moreover, industrialization and economic growth attracted many people to the regions and induced interventions in the environment, such as deforestation, landfill and irregular constructions (\cite{Londe2014, londe2015impactos}).

The susceptible flood areas used in this study were mapped by the Brazilian Geological Survey (CPRM), based on a database of previous occurrences and {\it in situ} evaluation of physical characteristics (\cite{CPRM2019}).

\subsection{Topological vulnerability}

Several topological measures can be extracted from a network and used to analyze the modeled phenomena or processes - see \cite{Costa2007}. One of these simple and important indexes is the shortest path length $d_{ij}$ between two nodes $i$ and $j$, defined as the smallest number of links from $i$ to $j$, among all the possible paths between $i$ and $j$. On the other hand, the efficiency $e_{ij}$ in the communication between nodes $i$ and $j$ can be defined as inversely proportional to shortest path length between them. The average efficiency $E$ of the network $G$ is defined as the average of all $e_{ij}$, considering all pairs of nodes. The topological vulnerability index of an element $k$ in a network $G$, $V_{k}$, is thus given by
\begin{equation}
V_{k} = \frac{E - E_{k}^{\star}}{E},
\end{equation}
where $E_{k}^{\star}$ is the efficiency of the network when the element $k$ is inaccessible: all its edges are removed. The first paper considering the pointwise vulnerability index was that of \cite{Goldshtein2004}, based on two relevant previous works: \cite{Latora2001, Latora2004}.

According to \cite{Pregnolato2016}, network models are typically aspatial: the emphasis has been on topological interactions, not on their geography. Mode details about space-related properties in Network Science can be found in \cite{Barth2011, Daqing2011}.

Here, we use the concept and tools of a (geo)graph, a network in a geographical space. Recently, this approach was applied for a mobility network analysis (\cite{Santos2019a}) and for a rainfall network analysis (\cite{Seron2019}). In this paper, we represent a set of highways as a network, calculate the topological vulnerability index of its elements and show them on a map. We highlight the spacial location of the most vulnerable element, in order to combine this information with the locations most susceptible to either floods or landslides.

\subsection{Systematic Literature Mapping}
\label{sec:slm}

In this section we present a systematic literature mapping of papers related to road networks, natural disasters and network vulnerability. We collected manuscripts from Google Scholar using the following search protocol: ``(graph OR network) AND (transport OR roads OR highways OR streets) AND (disaster OR flood OR flooding OR landslide) AND (vulnerability)''. The aim is to identify gaps and trends in literature.

We select the first forty papers and, as a cut-off criterion, we discard those whose abstracts and title deviates from the subject. We end up with twenty three articles grouped according to Table \ref{tab:slm}, in chronological publication order. 

\begin{landscape}

\begin{table}[]
\caption{\label{tab:slm}Results of the systematic literature mapping.}
\resizebox{1.35\textheight}{!}{%
\begin{tabular}{|l|
>{\columncolor[HTML]{FFFFFF}}r |
>{\columncolor[HTML]{FFFFFF}}l |
>{\columncolor[HTML]{FFFFFF}}l |
>{\columncolor[HTML]{FFFFFF}}l |
>{\columncolor[HTML]{FFFFFF}}l |
>{\columncolor[HTML]{FFFFFF}}l |
>{\columncolor[HTML]{FFFFFF}}l |}
\hline
\cellcolor[HTML]{FFFFFF}\textbf{Paper}                  & \multicolumn{1}{l|}{\cellcolor[HTML]{FFFFFF}\textbf{Citations}} & \textbf{Journal type}                        & \textbf{Country of the data} & \textbf{Context of ``vulnerability''} & \textbf{Network type}   & \textbf{Disaster type}              & \textbf{Metric ``vulnerability''}     \\ \hline
\cellcolor[HTML]{FFFFFF}\textit{\cite{Balijepalli2014}} & 119                                                             & Geography                                    & England                      & graph                                 & road network            & natural disaster                    & vulnerability                         \\ \hline
\cellcolor[HTML]{FFFFFF}\textit{\cite{Mattsson2015}}    & 605                                                             & Transport                                    & No Country                   & graph                                 & road system             & natural disaster                    & vulnerability                         \\ \hline
\cellcolor[HTML]{FFFFFF}\textit{\cite{Lu2016}}          & 12                                                              & interdisciplinary                            & China                        & transport                             & urban road network      & natural disaster                    & vulnerability                         \\ \hline
\cellcolor[HTML]{FFFFFF}\textit{\cite{Sköld2016}}       & 21                                                              & interdisciplinary                            & Swedish                      & disaster                              & streets                 & flooding                            & closeness                             \\ \hline
\cellcolor[HTML]{FFFFFF}\textit{\cite{Postance2017}}    & 45                                                              & interdisciplinary                            & United Kingdom               & transport                             & road network            & landslide                           & vulnerability                         \\ \hline
\cellcolor[HTML]{FFFFFF}\textit{\cite{Helderop2019}}    & 20                                                              & Transport                                    & United States                & transport                             & road network            & storm surge flooding                & {\color[HTML]{202124} susceptibility} \\ \hline
\cellcolor[HTML]{FFFFFF}\textit{\cite{Koks2019}}        & 159                                                             & interdisciplinary                            & No Country                   & transport                             & road/railway            & {\color[HTML]{202124} multi-hazard} & {\color[HTML]{202124} vulnerability}  \\ \hline
\cellcolor[HTML]{FFFFFF}\textit{\cite{Lu2019}}          & 17                                                              & interdisciplinary                            & China                        & graph                                 & urban rail transit      & vehicle breakdown                   & vulnerability                         \\ \hline
\cellcolor[HTML]{FFFFFF}\textit{\cite{Morales2019}}     & 9                                                               & interdisciplinary                            & Spain                        & transport                             & road network            & flooding                            & vulnerability                         \\ \hline
\cellcolor[HTML]{FFFFFF}\textit{\cite{Schlögl2019}}     & 25                                                              & interdisciplinary                            & Austrian                     & graph                                 & road network            & landslide                           & {\color[HTML]{202124} susceptibility} \\ \hline
\cellcolor[HTML]{FFFFFF}\textit{\cite{Abdulla2020}}     & 119                                                             & Transport Geography                          & England                      & graph                                 & road network            & flooding                            & vulnerability                         \\ \hline
\cellcolor[HTML]{FFFFFF}\textit{\cite{Fan2020}}         & {\color[HTML]{202124} 18}                                       & {\color[HTML]{202124} interdisciplinary}     & United States                & transport                             & road networks           & flooding                            & susceptibility                        \\ \hline
\cellcolor[HTML]{FFFFFF}\textit{\cite{Hearn2020}}       & 1                                                               & Geology                                      & Laos                         & disaster                              & road network            & landslide                           & vulnerability                         \\ \hline
\cellcolor[HTML]{FFFFFF}\textit{\cite{Jamshed2020}}     & 28                                                              & interdisciplinary                            & No Country                   & disaster                              & rural vulnerability     & flooding                            & vulnerability                         \\ \hline
\cellcolor[HTML]{FFFFFF}\textit{\cite{Mera2020}}        & 7                                                               & Transport                                    & United Kingdom               & transport                             & streets                 & natural disaster                    & vulnerability                         \\ \hline
\cellcolor[HTML]{FFFFFF}\textit{\cite{Wiśniewski2020}}  & 4                                                               & Transport                                    & Poland                       & transport                             & transport network urban & flooding                            & vulnerability                         \\ \hline
\cellcolor[HTML]{FFFFFF}\textit{\cite{Wang2020}}        & 9                                                               & interdisciplinary                            & China                        & transport                             & highway                 & {\color[HTML]{202124} flooding}     & {\color[HTML]{202124} vulnerability}  \\ \hline
\textit{\cite{Yin2020}}                                 & 8                                                               & interdisciplinary                            & China                        & disaster                              & highway                 & landslide                           & susceptibility                        \\ \hline
\textit{\cite{Zhang2020}}                               & 14                                                              & Transport                                    & United States                & graph                                 & transportation network  & landslide                           & vulnerability                         \\ \hline
\cellcolor[HTML]{FFFFFF}\textit{\cite{Alabbad2021}}     & 14                                                              & {\color[HTML]{202124} Environmental Science} & United States                & graph                                 & road network            & flooding                            & vulnerability                         \\ \hline
\cellcolor[HTML]{FFFFFF}\textit{\cite{Furno2021}}       & 1                                                               & interdisciplinary                            & France                       & graph                                 & transportation network  & traffic Jam                         & vulnerability                         \\ \hline
\textit{\cite{Morelli2021}}                             & 13                                                              & Transport                                    & Brazil                       & disaster                              & road network            & flooding                            & vulnerability                         \\ \hline
\textit{\cite{Shahdani2022}}                            & 0                                                               & interdisciplinary                            & Portugal                     & transport                             & road transport network  & flooding                            & vulnerability                         \\ \hline
\end{tabular}}
\end{table}

\end{landscape}

Regarding networks, Roads (highways) are the most frequent, followed by streets (urban networks). Nearly all papers quantify vulnerability with the network metric of the same name. The context of the word ``vulnerability'' relates more often to transport and graphs, with only a few associated to disasters.

In particular, we mention the top three cited papers: \cite{Mattsson2015} provide a review of problems in transport systems around the globe; \cite{Koks2019} bring a global view of multiple natural disasters; and \cite{Abdulla2020} investigate vulnerability of transport networks from British data. They all explore the problem under different scales: global and country-level.


Some future research directions: mapping of critical segments (roads) and the creation of susceptibility indexes to build catastrophe models in response to extreme events such as flooding and severe wind storms; mapping landslides incidence and; how severe and frequent the phenomena is becoming \citep{Postance2017}. The literature also lacks a detailed assessment of how climate transformations impact transport systems \citep{Morales2019}. 


Here we quantify links' vulnerability in a transportation network and their susceptibility to hazards and disruptions. We showed that the most well-cited works deal with a similar problem but at a country and global level, whereas we address it from a state-level perspective. 

\section{Results and Discussion}

Using the (geo)graph approach, we represent the set of highways as a network. For the Santa Catarina State case study area, the road network presents 1536 nodes/road segments and 2101 directed edges/connections between road segments.

\begin{figure}[!ht]
\includegraphics[width=14cm]{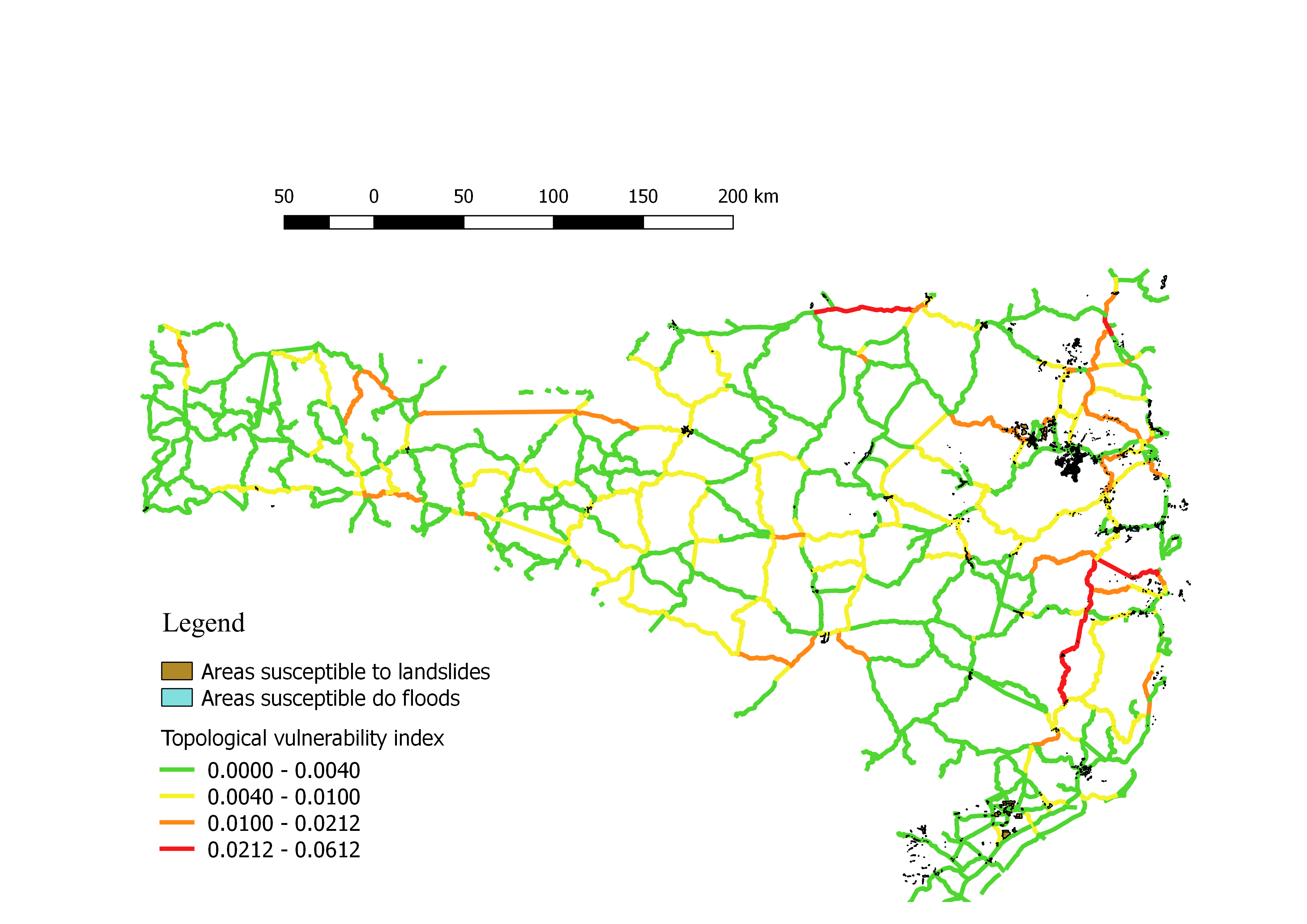}
\caption{Vulnerability index map for all highways in the study area. The green color is associated with the least vulnerable segments and the red color with the most vulnerable ones, following a quantile-scale color legend. Bounding-box: -53.73,-29.31,-48.54,-25.98 (geographical coordinates).
} \label{mapa1}
\end{figure}

Figure \ref{mapa1} shows the topological vulnerability index map for all highways in the study area. There are 4 classes (colors in Figure \ref{mapa1}), that corresponds to Low Vulnerability, Moderate Vulnerability, High Vulnerability, and Extremely High Vulnerability.

\begin{figure}[!ht]
\includegraphics[width=12cm]{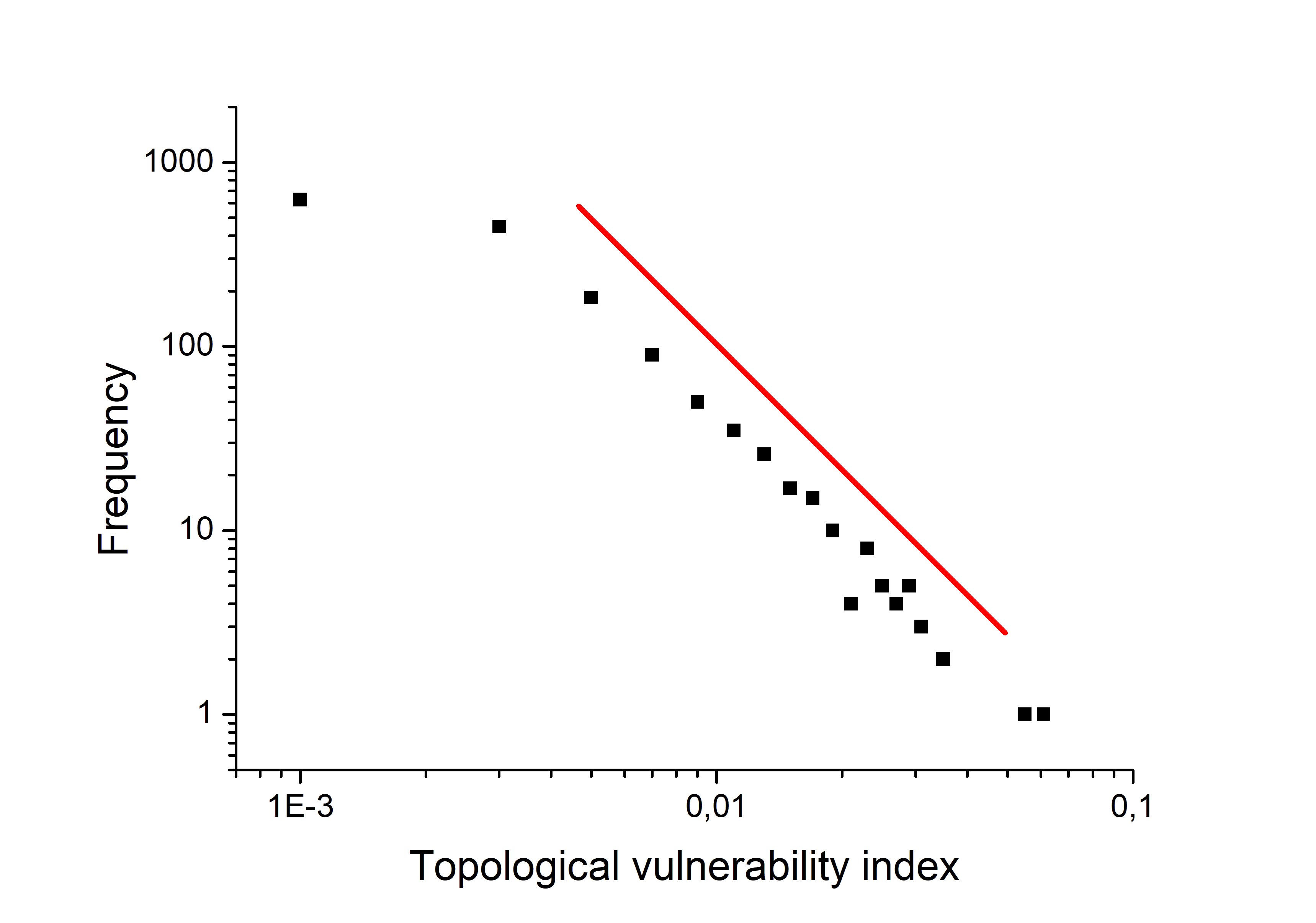}
\caption{Distribution of values for the topological vulnerability index (black squares) and a auxiliary line (red segment) for a power law with an exponent of $-2.475$).} \label{PL}
\end{figure}

The distribution of topological vulnerability index is highly inhomogeneous - see Figure \ref{PL}. In Figure \ref{PL}, on double logarithmic scale, we show this distribution and a power law fitting to it with an exponent as obtained by the Clauset's method (\cite{Clauset2009}). It means that most highways present low to moderate vulnerability, whereas a small subset is highly vulnerable.


Another important question is about where the most vulnerable elements are, in particular, whether they are close to the areas most susceptible to floods. The four most vulnerable segments are all on the SC-108 highway, including parts without pavement (in the rural area) and parts that cross areas susceptible to floods, in the urban area of Anitápolis/SC. In this city, there are several areas susceptible to floods and flash floods.

\begin{figure}[t]
\includegraphics[width=13cm]{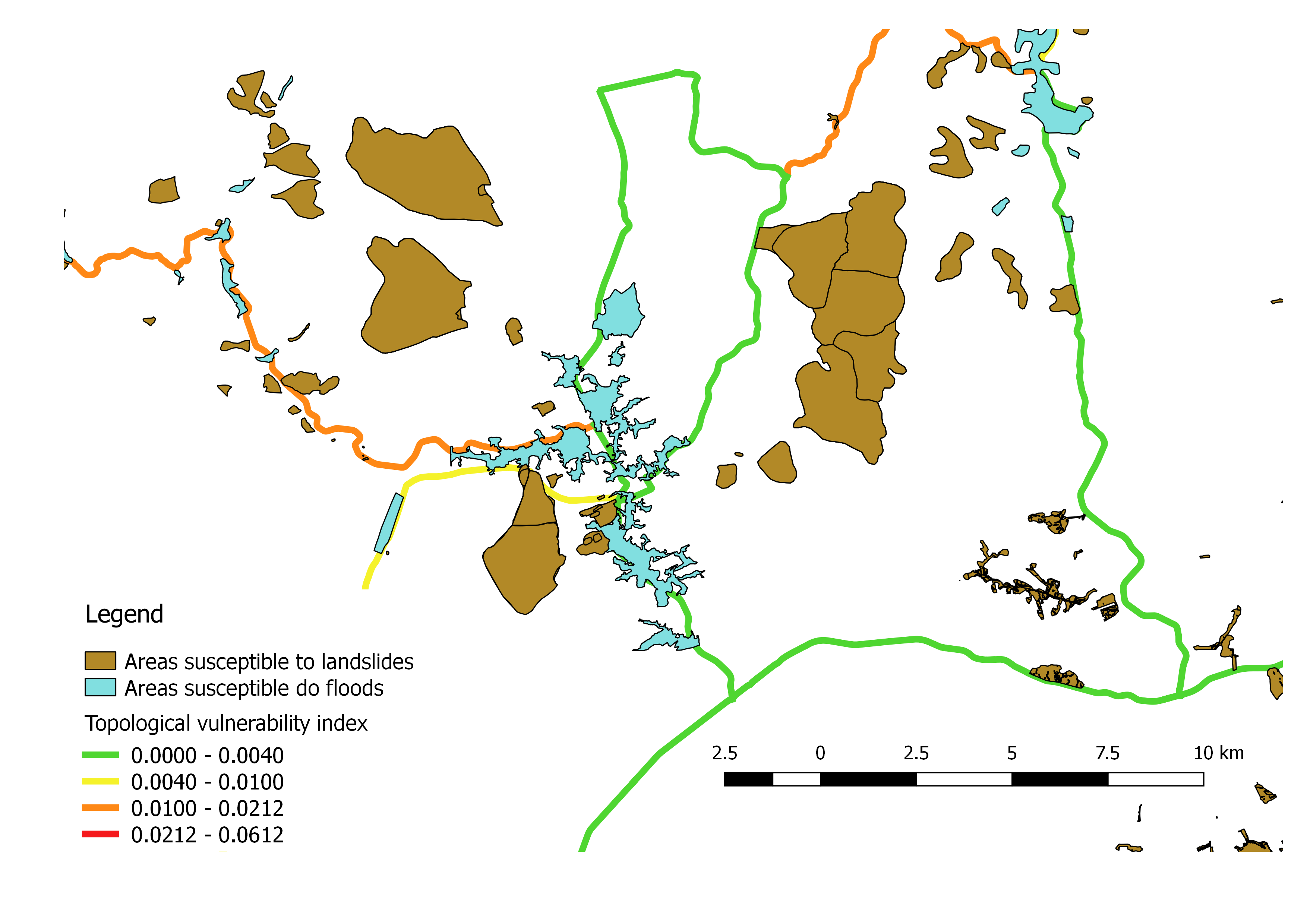}
\caption{Vulnerability index map for a subset of highways in the study area. The red color with the most vulnerable segments. The areas susceptible to flood are shown in blue.} \label{mapa2}
\end{figure}

Figure \ref{mapa2} shows the vulnerability index map for a subset of the highways in the study area, and, also, the areas most susceptible to hazards such as floods and landslides. In this subset, it is possible to see that there are some elements with high topological vulnerability index close to urban areas susceptible to flood. In this area, in the cities Rio Negrinho and Mafra, the BR-280 highway crosses the Negrinho River. This area is marked by several records of floods in the rainy season (susceptibility component), which makes traffic in the region unfeasible (impact) (UFSC, 2016).

The highway BR-280 is one of the most important in Santa Catarina state, playing an important role in the transportation of goods to the ports of São Francisco do Sul, Itajaí and Paranaguá. It also promotes the interconnection link between important cities in the region, such as Joinville and Jaraguá do Sul. Thus, there is a large flow of people and goods on this highway. 


This representation, considering both vulnerability (as a topological index) and susceptible areas, is an important tool for stakeholders from the transportation sector, considering climate change, disaster risk reduction and sustainable development agenda. The Risk Knowledge, combining hazard and vulnerability, is the first pillar of an Early Warning System (EWS) \cite{UNISDR2017}. Also, the transportation sector represents direct and indirect economic losses: the first being the destruction of physical assets and the second being a decline in economic value. In this work, the suggested representation ads knowledge about the losses to the disaster risk assessment.

\section{Conclusions}


In this paper, we represented the set of highways from our study area as a network and calculated the topological vulnerability index. Using the (geo) approach (\cite{Santos2019a, Seron2019}), it was possible to represent the results in a Geographical Information System.

In our case study, in the south region of Brazil, there are some elements with vulnerability index of approximately 5\%, therefore a flood impairing the traffic on this highway's element can reduce the efficiency of this transportation network by approximately 5\%. Also, there are elements high topological vulnerability index close to urban flood areas, for example, in the cities of Mafra and Rio Negrinho, where the BR-280 highway crosses the Negrinho River. This area is marked by several records of floods in the rainy season (susceptibility component), which makes traffic unfeasible in the region (impact). In the study area, the State of Santa Catarina, in Brazil, there is a heavy flow of people and goods, with some important national and international ports and airports.

The topological vulnerability index associated with an element of a network (in our case, a highway segment) is a measure quantifying the way the system reacts to damage on this element. Although it is a measure associated with the element, the topological vulnerability index contains information about the dynamics throughout the whole network (\cite{Santos2019b}). The disaster trigger is local but its impacts can be extended to a wider region. The topological vulnerability index captures this relation and it is possibly the most important feature of this index.

Accordingly to the Sendai Framework for Disaster Risk Reduction 2015-2030, one of the most important documents in the Disaster Risk Reduction (DRR) guidelines, there is one global target related to “Substantially reduce disaster damage to critical infrastructure and disruption of basic services” (\cite{Aitsi-Selmi2015}). Many critical infrastructures (such as roads) are of network type, and can be modeled using the Network Science approach (\cite{Newman2010}). The topological vulnerability index is a network measure particularly interesting in the context of critical infrastructures. The development of a vulnerability map to disasters and their impacts on infrastructures is aligned with the 2030 Agenda for Sustainable Development as well, particularly with the Sustainable Development Goals number 9, 11 and 13, related to Infrastructures, Intelligent Cities and Climate (\cite{SDG}).

To make better urban planning and to lessen the risk of disasters, mapping risk areas is an indispensable step. This mapping can be used to create a risk reduction plan, to define priority areas for attention in the municipalities, to make recommendations for works on infrastructure and to prepare municipal master plans. The mapping of risk areas for Santa Catarina follows the guidelines established from GIDES Project
(\cite{GIDES2018}), which is a partnership between Brazil and Japan to strengthen the National Strategy for integrated Management of Risks and Disasters. The project's goal is to reduce risks of disasters through non-structural preventive actions. The mains results are the improvement of assessment systems and risk mapping, warnings and urban planing for disaster prevention. 

A possible extension for this investigation is to draft risk scenarios considering other components, such as the dynamic exposure (daily traffic on each highway) and other kinds of vulnerability, for example, one based on traffic engineering parameters, or on the coverage of meteorological sensors (\cite{Carvalho2018}).

\section*{Author Contributions}

LBLS - Conceptualization, Methodology, Project Administration, Writing
GGS - Methodology, Writing
TG - Methodology, Writing
AASJ - Methodology, Writing
LRL - Conceptualization, Methodology, Writing
RTR - Methodology, Writing
RBB - Methodology, Writing
CESO - Methodology, Writing
VLSF - Methodology, Writing
IMS - Conceptualization, Review Editing

\section*{Acknowledgments}

Funding: S\~ao Paulo Research Foundation (FAPESP), Grant Number 2015/50122-0 and DFG-IRTG Grant Number 1740/2; FAPESP Grant Number 2018/06205-7; CNPq Grant Number 420338/2018-7

\section*{Data Availability Statement}
The datasets generated and analyzed for this study can be found in \url{https://github.com/gioguarnieri/highways-vulnerability}

\nolinenumbers

\bibliographystyle{apalike}
\bibliography{references}

\newpage

\end{document}